\documentclass[twocolumn,aps,prl,psfig,showpacs]{revtex4}
\usepackage{amsfonts}
\usepackage{amsmath}
\usepackage{graphicx}
\usepackage{bm}
\usepackage{amssymb}
\usepackage{times}
\usepackage{dcolumn}
\usepackage{psfig}

\makeatletter

\begin{document}

\title{Crystalline and Electronic Structures of Molecular Solid C$_{50}$Cl$%
_{10}$: First-Principles Calculation}
\author{Qing-Bo Yan}
\author{Qing-Rong Zheng}
\author{Gang Su}
\email[Author to whom correspondence should be addressed. ]{Email:
gsu@gucas.ac.cn}
\affiliation{{College of Physical Sciences, Graduate University of Chinese Academy of
Sciences, P.O. Box 4588, Beijing 100049, China}}

\begin{abstract}
A molecular solid C$_{50}$Cl$_{10}$ with possible crystalline structures,
including the hexagonal-close-packed (hcp) phase, the face-centered cubic
(fcc) phase, and a hexagonal monolayer, is predicted in terms of
first-principles calculation within the density functional theory. The
stable structures are determined from the total-energy calculations, where
the hcp phase is uncovered more stable than the fcc phase and the hexagonal
monolayer in energy per molecule. The energy bands and density of states for
hcp and fcc C$_{50}$Cl$_{10}$ are presented. The results show that C$_{50}$Cl%
$_{10}$ molecules can form either a hcp or fcc indirect-gap band insulator
or an insulating hexagonal monolayer.
\end{abstract}

\maketitle

Fullerenes such as C$_{60}$, C$_{70}$ and their larger homologs satisfy the
so-called isolated pentagon rule (IPR)\cite{Kroto_IPR,IPR_book}. This rule
states that the most stable fullerenes are those in which every pentagon is
surrounded by five hexagons. For some time, people believe that the non-IPR
fullerenes could be unstable due to their adjacent pentagons, and may be
difficult to synthesize. In recent years, it has been uncovered, however,
that the IPR can be violated by metallic endohedral or exohedral chemical
deriving fullerenes\cite{endohedral_CRWang, endohedral_Stevenson,
endohedral_Kato,exohedral_Hummelen, exohedral_Nuber_Hirsch,
exohedral_Nuber_Keshavarzk}. On account of the adjacent pentagons and the
high curvature of molecular surface, the non-IPR fullerenes might have
unusual electronic, magnetic, and mechanical properties. Last year, Xie 
\textit{et al.}\cite{Xie} had successfully synthesized a new exohedral
chemical deriving non-IPR D$_{5h}$ fullerene[50], say, C$_{50}$Cl$_{10}$,
with 10 chlorine atoms added to the pentagon-pentagon vertex fusions. The
molecular structure of C$_{50}$Cl$_{10}$ looks like a Saturn-shaped profile,
as depicted in Fig. 1, which is compatible with the Euler theorem, ensuring
that the fullerenes with fewer than sixty C atoms and comprised of only
pentagons and hexagons do not satisfy the IPR. The existence of D$_{5h}$ C$%
_{50}$Cl$_{10}$ is evidenced by mass spectrum, and $^{13}$C NMR spectra; and
of particular interest is that C$_{50}$Cl$_{10}$ could easily react with a
variety of organic groups to form new compounds\cite{Xie} which might show
interesting chemical and physical properties. Since its discovery, there
have been several investigations\cite%
{c50cl10peapod,c50cl10polarizability,c50cl10optical,c50cl10luxin_jacs,c50cl10buildingblock_jpc,c50sphericity_cpl}
on the electronic and optical properties of C$_{50}$Cl$_{10}$ clusters.

So far, none has discussed, however, the possibility if C$_{50}$Cl$_{10}$
molecules can form a solid. As is well-known, C$_{60}$ clusters can condense
to form a solid with either the face-centered-cubic (fcc) phase\cite{fleming}
or the hexagonal-close-packed (hcp) phase\cite{kratschmer}. Although the
molecular structure is different, there is no reason to believe that C$_{50}$%
Cl$_{10}$ molecules cannot condense to form a solid. Therefore, it would be
interesting to address this issue at least from a viewpoint of numerical
simulations. In this paper, we report a first-principles study on the
possible solid C$_{50}$Cl$_{10}$ within the framework of density functional
theory (DFT) \cite{Hohenberg-Kohn} with local density approximation (LDA)%
\cite{Kohn-Sham}. From the total-energy calculations we have found that C$%
_{50}$Cl$_{10}$ molecules could form a stable hcp or fcc crystal or a
hexagonal monolayer, though the hcp phase is more stable than the fcc phase
and the hexagonal monolayer in energy per molecule. The optimized lattice
constants are $a_{0}=13.57$ $\mathring{A}$, $c_{0}=11.39$ $\mathring{A}$ for
the hcp phase, $a_{0}=19.20$ $\mathring{A}$ for the fcc phase, and $%
a_{0}=13.52$ $\mathring{A}$ for the hexagonal monolayer. The energy bands
and density of states (DOS) for three structures have been calculated. The
results show that the solid C$_{50}$Cl$_{10}$ is a band insulator with an
indirect energy gap of $1.59$ $eV$ for the hcp phase or $1.79$ $eV$ for the
fcc phase, and with a gap of $1.85$ $eV$ for the hexagonal monolayer, where
the gap widths are comparable with those of C$_{60}$ solids\cite%
{cohesive_solid_C60, c60solid_gap}.

\begin{figure}[tbp]
\includegraphics[width=0.85\linewidth,clip]{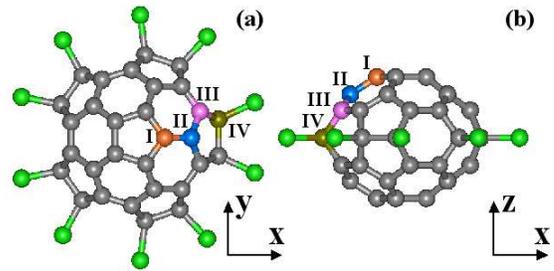}
\caption{(Color online) Schematic structure of molecule C$_{50}$Cl$_{10}$:
(a) top view, and (b) side view. The dark grey and green (light grey) balls
represent C and Cl atoms, respectively. The numbers I to IV indicate the
four types of C atoms.}
\label{c50cl10_structure}
\end{figure}

The total energies, energy bands and DOS are calculated by means of the
ABINIT package\cite{ABINIT}. This package is coded within the DFT framework
based on pseudopotentials and plane waves, which relies on an efficient fast
Fourier transform algorithm\cite{FFT} for the conversion of wave functions
between real and reciprocal spaces, on the adaptation to a fixed potential
of the band-by-band conjugate gradient method\cite{Conjugate Gradient}, and
on a potential-based conjugate-gradient algorithm for the determination of
the self-consistent potential \cite{Gonze1,Gonze2,Gonze3}. Troullier-Martins
norm conserving pseudopotentials \cite{pseudo potential} are applied to
mimic the electron-ion interaction, and the Teter parametrization\cite{Teter}
of the Ceperley-Alder exchange-correlation potential is used. The kinetic
energy cutoff in the plane-wave basis is taken as 20 Hartree, and the
tolerance for absolute differences of the total energy is set as 10$^{-6}$
Hartree. The convergence of the total energy to the kinetic energy cutoff
has been checked.

As shown in Fig. 1, the C$_{50}$Cl$_{10}$ molecule has 4 unique types of
carbon atoms (\textnormal{I} to \textnormal{IV}) within D$_{5h}$ symmetry,
and ten Cl atoms are added to the most reactive C$_{\mathnormal{IV}}$ sites
(i.e. to pentagon-pentagon vertex fusions). Such a structure obviously
violates the IPR. The C-Cl bond length in C$_{50}$Cl$_{10}$ is 1.80 ${%
\mathnormal{\mathring{A}}}$\cite{Xie}, which is comparable with that in CCl$%
_{4}$, 1.78 ${\mathnormal{\mathring{A}}}$. We have calculated the
valence-electron density of C$_{50}$Cl$_{10}$, and found that the high
density of the C-Cl bond shows that the ten Cl atoms are indeed bonded with
the C$_{50}$ cage, and a stable C$_{50}$Cl$_{10}$ molecule can be formed.

\begin{figure}[tbp]
\includegraphics[width=3.0in, keepaspectratio]{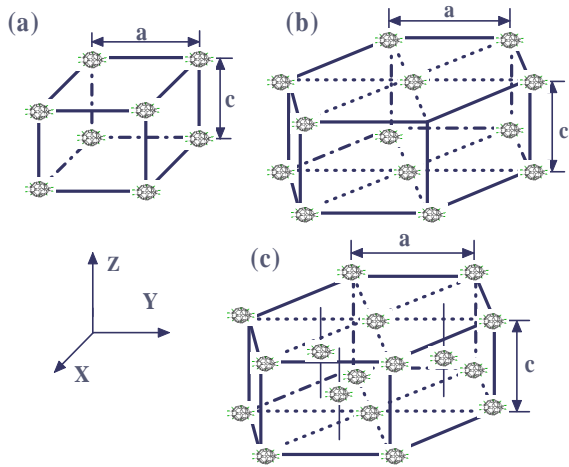}
\caption{The presumed crystalline structures: (a) a tetragonal lattice; (b)
a simple hexagonal lattice; and (c) a hcp lattice. Note that to show
lattices clearly, the C$_{50}$Cl$_{10}$ molecules are drawn much smaller
than those in the real proportion.}
\label{c50cl10_lattice_mol}
\end{figure}

To seek for stable solid phases, by carefully analyzing the geometrical
symmetry of C$_{50}$Cl$_{10}$ where it has the D$_{5h}$ symmetry in $xy$
plane with diameter (10.3 ${\mathnormal{\mathring{A}}}$) almost twice larger
than that along $z$ axis (5.4 ${\mathnormal{\mathring{A}}}$), we find the
crystal lattices such as tetragonal (tt), simple hexagonal (sh), fcc and hcp
structures could be the most possible candidates. The former three are shown
in Fig. 2, where the coordination systems of the lattices are coincident
with those of the C$_{50}$Cl$_{10}$ molecule. Then, the total energy is
calculated accordingly with respect to different lattice parameters for each
presumed structure\cite{note1}. The total energy is scanned in every 0.1 $%
\mathnormal{\mathring{A}}$ for lattice constants $a$ and $c$. Owing to the
limits of the method and the capability of our computer system, a properly
fixed orientation of C$_{50}$Cl$_{10}$ molecules, namely, the $xy$ plane of C%
$_{50} $Cl$_{10}$ molecule is arranged to be vertical to the $z$ axis in tt,
sh, and hcp structures, and to be along the (111) plane in the fcc
structure, is assumed in calculations, which makes the system bear a
relatively high symmetry\cite{note2}. 
\begin{table}[b]
\caption{The calculated minimal relative total energies per C$_{50}$Cl$_{10}$
molecule and the corresponding lattice constants for the presumed four solid
structures.}\setlength{\textfloatsep}{0.1cm} 
\begin{tabular}{ccccc}
\hline\hline
&  & \multicolumn{2}{c}{Lattice constants (${\mathnormal{\mathring{A}}}$)} & 
\\ \cline{3-4}
& Minimums & \hspace{7mm}$a$\hspace{7mm} & $c$ & Energy (eV) \\ \hline
& A & $13.99$ & $9.22$ & $0.77$ \\ \cline{2-5}
& B & $13.61$ & $9.72$ & $0.77$ \\ \cline{2-5}
tt & C & $14.02$ & $9.80$ & $0.76$ \\ \cline{2-5}
& D & $13.72$ & $10.31$ & $0.75$ \\ \cline{2-5}
& E & $13.99$ & $10.82$ & $0.73$ \\ \cline{2-5}
& F & $13.65$ & $11.41$ & $0.74$ \\ \hline
& A & $13.37$ & $9.75$ & $0.55$ \\ \cline{2-5}
& B & $13.38$ & $10.42$ & $0.58$ \\ \cline{2-5}
sh & C & $13.33$ & $10.79$ & $0.57$ \\ \cline{2-5}
& D & $13.39$ & $11.36$ & $0.54$ \\ \cline{2-5}
& E & $13.38$ & $12.37$ & $0.54$ \\ \cline{2-5}
& F & $13.38$ & $12.99$ & $0.55$ \\ \hline
hcp & A & $13.57$ & $11.39$ & $0.00$ \\ \hline
fcc & A & $19.20$ & $-$ & $0.60$ \\ \hline\hline
\end{tabular}%
\end{table}

Table I gives the calculated minimal relative total energies per C$_{50}$Cl$%
_{10}$ molecule and the corresponding lattice constants for the four
presumed solid phases, where the data with precision of 0.01 $\mathring{A}$
is obtained by a bicubic interpolation from the calculated data with
precision of 0.1 $\mathring{A}$. For both tt and sh structures, the total
energy shows a series of minimums (labeled by A-F) with almost the same
values, suggesting that these two structures are labile. For the hcp and fcc
structures, only one minimal total energy is obtained, showing that the hcp
and fcc C$_{50}$Cl$_{10}$ could be formed, though the hcp phase looks more
stable than the fcc phase in energy per molecule, as the minimal energy of
the hcp structure is about 0.6 $eV$ per C$_{50}$Cl$_{10}$ molecule lower
than those of tt, sh and fcc structures.

\begin{figure}[bp]
\includegraphics[width=0.85\linewidth,clip]{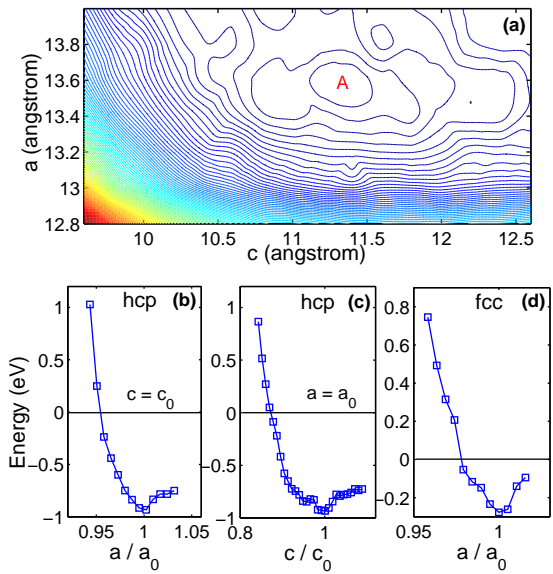}
\caption{(Color online) (a) The contour map of total energy for hcp C$_{50}$%
Cl$_{10}$ in the plane of lattice constants $a$ and $c$, where A indicates
the minimum of the total energy. The total energy per C$_{50}$Cl$_{10}$
molecule, measured from the total energy of the isolated C$_{50}$Cl$_{10}$
molecule, as a function of the lattice constant $a$ (b) and $c$ (c) for hcp
phase, where the optimized lattice constants $a_{0}=$ 13.57 ${\mathnormal{%
\mathring{A}}}$, $c_{0}=$ 11.39 ${\mathnormal{\mathring{A}}}$, and (d) for
fcc phase with the optimized lattice constant $a_{0}=$ 19.20 ${\mathnormal{%
\mathring{A}}}$.}
\label{c50cl10_hcp_contour_curve}
\end{figure}

To show it clearly, the contour map of the total energy\emph{\ }for the hcp C%
$_{50}$Cl$_{10}$ in the plane of lattice constants $a$ and $c$ is depicted
in Fig. 3(a), where only one minimum of the total energy is observed at $%
a=a_{0}=$13.57 ${\mathnormal{\mathring{A}}}$, and $c=c_{0}=$11.39 ${%
\mathnormal{\mathring{A}}}$, as marked by the letter A. Figs. 3(b) and (c)
present the total energy per C$_{50}$Cl$_{10}$ molecule of the hcp C$_{50}$Cl%
$_{10}$ crystal, which is measured from the total energy of isolated C$_{50}$%
Cl$_{10}$ clusters\cite{note3}, as a function of lattice constants $a$ and $%
c $, respectively. The results reveal that the hcp structure is stable. Fig.
3(d) gives the total energy per C$_{50}$Cl$_{10}$ cluster in the fcc C$_{50}$%
Cl$_{10}$ crystal as a function of lattice constant $a$, where only one
minimum of the total energy is seen at $a=a_{0}=$19.20 ${\mathnormal{%
\mathring{A}}}$, indicating that the fcc C$_{50}$Cl$_{10}$ crystal could
also be stable, though the minimal total energy of the fcc phase is larger
than that of the hcp phase.

In order to confirm the relative stabilities of the obtained hcp and fcc
lattices, we have optimized the atomic positions with fixed lattice
parameters by 40 steps Broyden-Fletcher-Goldfarb-Shanno minimization. The
results show that no notable atomic displacements are observed, and the
total-energy changes are less than 0.0003\% (0.04eV), indicating that the C$%
_{50}$Cl$_{10}$ molecules keep their shapes and sizes, and the lattice
structures which we have obtained are stable.

The calculated cohesive energy per C$_{50}$Cl$_{10}$ molecule is 0.9 $eV$
for the hcp solid, and 0.3 $eV$ for the fcc solid, which is much smaller
than the typical C-C bond energy (more than 3 $eV$). The cohesive energies
of both solid phases are less than that of fcc C$_{60}$ solid (1.6 $eV$) 
\cite{cohesive_solid_C60}, implying that C$_{50}$Cl$_{10}$ solid is less
stable than C$_{60}$ solid. The contour map of the valence-electron density
for hcp C$_{50}$Cl$_{10}$ crystal is given in Fig. 4. It is seen that the
valence-electron density between C$_{50}$Cl$_{10}$ molecules is much lower
than the density in a single C$_{50}$Cl$_{10}$ molecule. The similar
situation occurs for the fcc phase of C$_{50}$Cl$_{10}$.

\begin{figure}[tbp]
\includegraphics[width=2.0in, keepaspectratio]{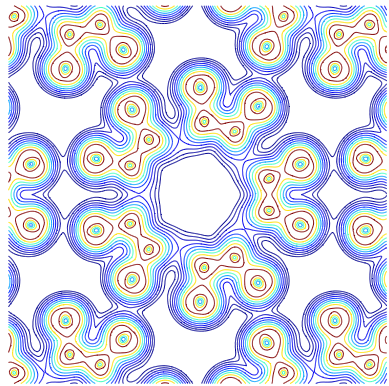}
\caption{(Color online) Contour map of the valence-electron density of hcp C$%
_{50}$Cl$_{10}$ crystal in the (001) plane. The larger lobe-like patterns
stand for those of Cl atoms, while the inner smaller dumbbell-like patterns
stand for those of C atoms, where the highest-density contour (red, around C
and Cl atoms) represents the equal-density line of 1.383 /$\mathring{A}^{3}$%
, and the lowest-density contour (blue, dark grey) represents the
equal-density line of 0.054 /$\mathring{A}^{3}$. It shows that the
interactions between the Cl atoms in the two neighboring C$_{50}$Cl$_{10}$
molecules are quite weak.}
\end{figure}

The energy bands, DOS $G(\varepsilon )$, as well as the integrated DOS $%
I(E)=\int_{-\infty }^{E}G(\varepsilon )d{\varepsilon }$ for solid C$_{50}$Cl$%
_{10}$ are calculated for the hcp and fcc phases with the optimized lattice
constants $a_{0}$ and $c_{0}$. Fig. 5 presents the band structures of the
hcp and fcc C$_{50}$Cl$_{10}$ crystals, respectively, where only several
energy bands around the energy gap are shown. For the hcp crystal, the
energy difference between the valence-band top (VBT) at the $\Gamma $ point
and the conduction-band bottom (CBB) at the $K$ point is 1.59 $eV$, and for
the fcc crystal, the energy gap between the VBT at the $\Gamma $ point and
the CBB at the $L$ point is about 1.79 $eV$, showing that both hcp and fcc C$%
_{50}$Cl$_{10}$ solids may be indirect-gap band insulators. The present
situation is in contrast to the fcc C$_{60}$ solid which is a direct-gap
insulator with an energy gap of 1.5 $eV$\cite{cohesive_solid_C60}. We have
also calculated the energy levels of C$_{50}$Cl$_{10}$ molecule, and found
that the energy gap between the highest-occupied molecular orbital (HOMO)
and the lowest-unoccupied molecular orbital (LUMO) is around 1.9 $eV$, which
is in well agreement with a previous calculated result\cite%
{c50cl10polarizability}, and is comparable with that of C$_{60}$ cluster
where the energy gap between the HOMO state, $h_{u}$, and the LUMO state, $%
t_{1u}$, is about 1.9 $eV$\cite{cohesive_solid_C60}. Obviously, the energy
gap remains finite when the C$_{50}$Cl$_{10}$ molecules are condensed to
form a solid. Like other typical molecular systems\cite{c60solid_narrowbands}%
, the energy bands of these two C$_{50}$Cl$_{10}$ crystals become narrow, as
they are formed through the overlap of C$_{50}$Cl$_{10}$ molecular energy
levels.

\begin{figure}[tbp]
\includegraphics[width=4.0in, keepaspectratio]{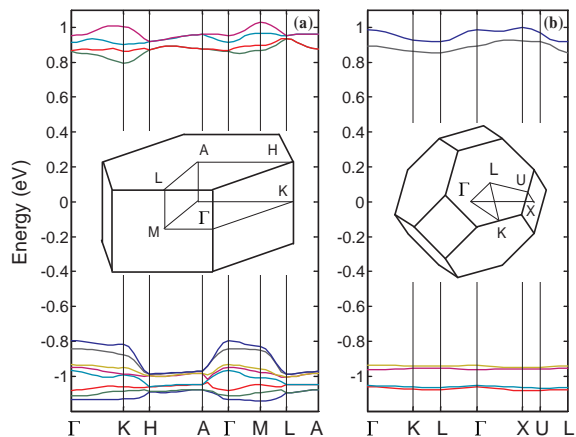}
\caption{(Color online) Band structures of hcp (a) and fcc (b) C$_{50}$Cl$%
_{10}$ around the energy gap, where the Fermi level is set to zero. For the
hcp phase, the valence-band top (VBT) is at $\Gamma $ point, and the
conduction-band bottom (CBB) is at $\mathnormal{K}$ point; for the fcc
phase, the VBT is at the $\Gamma $ point, and the CBB is at the $L$ point,
showing the solid C$_{50}$Cl$_{10}$ is an indirect band insulator with a gap
of 1.59 $eV$ (hcp) or 1.79 $eV$ (fcc).}
\label{c50cl10_hcp_band}
\end{figure}

\begin{figure}[bp]
\includegraphics[width=3.0in,keepaspectratio]{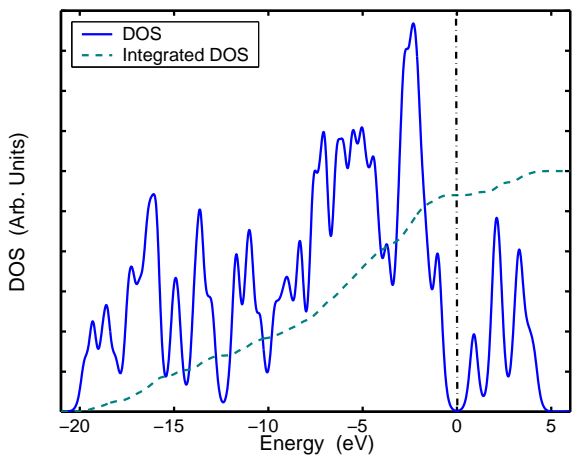}
\caption{(Color online) Density of states (DOS) and the integrated DOS
(dashed line) of electrons as a function of energy for the hcp C$_{50}$Cl$%
_{10}$. The energy zero-point is taken at the Fermi level, that is indicated
by the vertical dash-dot line. The flat part of the integrated DOS in the
vicinity of the Fermi level gives the width of energy gap.}
\label{c50cl10_hcp_dos}
\end{figure}

Fig. 6 presents the DOS and the integrated DOS of the hcp C$_{50}$Cl$_{10}$
as an example, which are obtained by means of a smearing technique. An
energy gap in the vicinity of the Fermi level is clearly seen. From the
positions of peaks in the DOS around the energy gap, the information of the
HOMO-derived and LUMO-derived bands as well as next higher bands can be
extracted. Evidently, the DOS of electrons at the LUMO-derived states
appears to be smaller than that at the HOMO-derived states. With the
increase of energy, the integrated DOS increases nonmonotonically below and
above the energy gap, where the flat part in the vicinity of the Fermi level
measures the width of the energy gap. The presented results in Fig. 6 are in
agreement with the calculated electronic energy levels shown in Fig. 5. It
should be remarked that the obtained DOS as a function of energy shows a
very similar behavior for the hcp and fcc phases, though there are slightly
quantitative changes. The reason is that owing to the quite weak
interactions between C$_{50}$Cl$_{10}$ molecules, as manifested in Fig. 4,
the DOS of electrons for solids would have profiles similar to that for
clusters, yielding that the DOS is similar for both hcp and fcc C$_{50}$Cl$%
_{10}$.

Let us now address another interesting issue if C$_{50}$Cl$_{10}$ molecules
can condense to form a monolayer structure. Recall that C$_{60}$ clusters
can form a hexagonal monolayer structure which has gained much attention
recently\cite{C60-monolayer}. From Table I, one may note that for the sh
lattice, the minimal total energies have slight changes with the lattice
constant $a$, but have remarkable changes with constant $c$. It hints that a
stable layered structure with a single minimal total energy could probably
be formed. To probe this possibility, we have taken the lattice constant $c$
of the sh supercell to 40 ${\mathnormal{\mathring{A}}}$ to decline the
interactions between the C$_{50}$Cl$_{10}$ layers but keeping the molecular
orientation the same as in the sh lattice. It turns out, after some efforts,
that a single minimal total energy with the optimized lattice constant $%
a_{0}=13.52$ ${\mathnormal{\mathring{A}}}$ is indeed obtained. The cohesive
energy per C$_{50}$Cl$_{10}$ molecule for this monolayer is found to be 0.3 $%
eV$. The energy bands and DOS show that the C$_{50}$Cl$_{10}$ hexagonal
monolayer is a band insulator with an energy gap of 1.85 $eV$.

It would be emphasized that as manifested in Fig. 4, the molecule-molecule
interactions in the solid C$_{50}$Cl$_{10}$ appear to be weak. Whether the
interactions between C$_{50}$Cl$_{10}$ molecules in the solid phase are of
the van der Waals force or not, is at present unclear within our
calculations, because it is known that the LDA cannot correctly describe the
van der Waals force as it assumes fundamentally that the electron density is
flat like in metals (see, e.g. \cite{Wu}). Nevertheless, our calculations
indeed suggest that the molecular solid C$_{50}$Cl$_{10}$ with either the
hcp, fcc or hexagonal monolayer phase can be somehow formed through weak
interactions.

To summarize, in terms of first-principles calculation within the density
functional theory, we have predicted a molecular solid C$_{50}$Cl$_{10}$
with either a hcp, fcc, or a hexagonal monolayer structure. These possibly
stable structures are determined from the total-energy calculations, where
the hcp phase is found more stable than the fcc phase and the hexagonal
monolayer in energy per molecule. The energy bands and DOS for the hcp and
fcc C$_{50}$Cl$_{10}$ as well as the hexagonal monolayer are discussed. Our
results show that C$_{50}$Cl$_{10}$ solid can be a band insulator with an
indirect energy gap of 1.59 $eV$ \ for the hcp structure and 1.79 $eV$ for
the fcc structure, or with a gap of 1.85 $eV$ for the hexagonal monolayer.
Since the LDA calculation usually underestimates the energy gap \cite%
{LDA_gap,c60solid_gap}, our present simulations inevitably call for further
theoretical studies \cite{note4} or experiments. We expect that the C$_{50}$%
Cl$_{10}$ solids can be successfully synthesized in near future.

The authors are grateful to B. Gu, H. F. Mu, X. B. Wang, and Z. C. Wang for
helpful discussions. This work is supported in part by the National Science
Foundation of China (Grant Nos. 90403036, 20490210), and by the MOST of
China (Grant No. 2006CB601102).


\begin{thebibliography}{99}
\bibitem{note1} For the tetragonal, simple hexagonal, and hcp structures, as
we have two lattice constants $a$ (in $xy$ plane) and $c$ (along $z$ axis),
the total energy is a bivar function of $a$ and $c$. For the fcc structure,
the total energy is a function of lattice constant $a$.
\bibitem{note2} After we have determined the proper lattice structures and
parameters, we calculated the change of the total energy by rotating C$_{50}$%
Cl$_{10}$ molecules around $z$ axis. It has been found that the change is no
more than 0.0008\% (0.09eV), suggesting that the effect of molecular
rotations on the total energy could be ignored. As the C$_{50}$Cl$_{10}$
diameter in $xy$ plane is almost twice larger than that along $z$ axis,
being nearly the lattice parameter along $z$ axis in the hcp structure, the
rotation around $x $ or $y$ axis will break the lattice structure, and would
not be possible. Similar situations occur for the other interested lattice
structures.
\bibitem{note3} Note that the isolated C$_{50}$Cl$_{10}$\ molecule has been
simulated, as usual, by a cubic supercell calculation with a lattice
parameter 26 $\mathring{A}$.
\bibitem{note4} As the relative stability of different structures of
molecular crystals would not be well determined by LDA, an independent
confirmation by other methods for the structures predicted in the present
paper is called on.
\end{thebibliography}
\end{document}